\documentclass[prb,superscriptaddress,preprint, floats]{revtex4}
\usepackage{graphicx}
\usepackage{bm}
\usepackage{pifont}
\usepackage{amsmath}
\usepackage{color}

\newcommand {\SRO}{Sr$_2$RuO$_4$}
\newcommand {\Tc}{$T_{\mathrm{c}}$}
\newcommand {\Rn}{$R_{\mathrm{n}}$}
\newcommand {\Ic}{$I_{\mathrm{c}}$}
\newcommand {\Icp}{$I_{\mathrm{c,p}}$}
\newcommand {\Icn}{$I_{\mathrm{c,n}}$}

\newcommand {\IcRn}{$I_{\mathrm{c}}R_{\mathrm{n}}$}

\newcommand {\pxyp}{$p_{x}+ ip_{y}$}
\newcommand {\pxym}{$p_{x}- ip_{y}$}
\newcommand {\dxsys}{$d_{x^2-y^2}$}

\begin{document}
\title{Characterization of Sr$_2$RuO$_4$ Josephson junctions\\
made of epitaxial films} 
\author{Masaki Uchida}
\email[Author to whom correspondence should be addressed: ]{uchida@ap.t.u-tokyo.ac.jp}
\affiliation{Department of Applied Physics and Quantum-Phase Electronics Center (QPEC), the University of Tokyo, Tokyo 113-8656, Japan}
\affiliation{PRESTO, Japan Science and Technology Agency (JST), Tokyo 102-0076, Japan}
\author{Ikkei Sakuraba}
\affiliation{Department of Applied Physics and Quantum-Phase Electronics Center (QPEC), the University of Tokyo, Tokyo 113-8656, Japan}
\author{Minoru Kawamura}
\affiliation{RIKEN Center for Emergent Matter Science (CEMS), Wako 351-0198, Japan}
\author{Motoharu Ide}
\affiliation{Department of Applied Physics and Quantum-Phase Electronics Center (QPEC), the University of Tokyo, Tokyo 113-8656, Japan}
\author{Kei S. Takahashi}
\affiliation{PRESTO, Japan Science and Technology Agency (JST), Tokyo 102-0076, Japan}
\affiliation{RIKEN Center for Emergent Matter Science (CEMS), Wako 351-0198, Japan}
\author{Yoshinori Tokura}
\affiliation{Department of Applied Physics and Quantum-Phase Electronics Center (QPEC), the University of Tokyo, Tokyo 113-8656, Japan}
\affiliation{RIKEN Center for Emergent Matter Science (CEMS), Wako 351-0198, Japan}
\author{Masashi Kawasaki}
\affiliation{Department of Applied Physics and Quantum-Phase Electronics Center (QPEC), the University of Tokyo, Tokyo 113-8656, Japan}
\affiliation{RIKEN Center for Emergent Matter Science (CEMS), Wako 351-0198, Japan}

\begin{abstract}
We have studied fundamental properties of weak-link {\SRO}/{\SRO} Josephson junctions fabricated by making a narrow constriction on superconducting {\SRO} films through laser micro-patterning. The junctions show a typical overdamped behavior with much higher critical current density, compared with those previously reported for bulk {\SRO}/$s$-wave superconductor junctions. Observed magnetic field and temperature dependences of the Josephson critical current suggest that the chiral $p$-wave is unlikely for the superconducting symmetry, encouraging further theoretical calculations of the {\SRO}/{\SRO} type junctions.

\end{abstract}
\maketitle

The discovery of superconductivity in {\SRO} \cite{SRO} has been followed by many important experimental and theoretical studies on unconventional superconductivity and its underlying physics, especially in terms of a possible chiral $p$-wave order parameter symmetry \cite{SROreview1, SROreview3desiringfilm, SROreview4desiringfilm}. In the chiral $p$-wave superconducting state, a finite orbital angular momentum of the paired electrons spontaneously breaks the time-reversal symmetry. The spin-triplet paring has been supported mainly by nuclear magnetic resonance (NMR) spectroscopy, observing unchanged Knight shift or spin susceptibility through the transition temperature {\Tc} \cite{SRO_NMR1}. The broken time-reversal symmetry has been indicated by muon spin relaxation ($\mu$SR)  \cite{SRO_usR} and magneto-optical Kerr effect (MOKE) \cite{SRO_MOKE}. On the other hand, there are other fundamental observations irreconcilable with the chiral $p$-wave state. Suppression of the upper critical field, which may be caused by the paramagnetic pair breaking for the in-plane magnetic field, is a typical example \cite{Hc2_yonezawafirst1, Hc2_thinfilm, SROreview4desiringfilm}.  In addition, chiral currents, emerging on the edge of chiral domains, have not been directly detected even by scanning superconducting quantum interference device (SQUID) microscopy with ultrahigh sensitivity \cite{SRO_nochiralcurrent1, SRO_nochiralcurrent2}.  While this result could be explained by mixed chiral domains with small sizes ($< 2$ $\mu$m), the assumed domain sizes are substantially different even between the $\mu$SR ($< 2$ $\mu$m) and MOKE ($> 50$ $\mu$m) experiments \cite{TRSB}. Recent thermal conductivity and specific heat measurements also favor other $d$-wave order parameters with line nodes \cite{SRO_specificheat, SRO_thermalconductivity}. Furthermore, a pronounced drop of the in-plane spin susceptibility has been recently confirmed by lower excitation power NMR experiments, which particularly rules out the possibility of the chiral $p$-wave state \cite{SRO_NMR2}. In this context, development of {\SRO} film-based junctions has been increasingly demanded for examining its superconducting state \cite{SROreview3desiringfilm, SROreview4desiringfilm}.

Phase sensitive experiments are capable of providing clear evidences for elucidating the superconducting pairing symmetry \cite{textbook1}. In the case of high-{\Tc} cuprates, for example, Josephson junctions fabricated from their epitaxial films have lead the determination of {\dxsys}-wave order parameter symmetry. On the other hand, successful growth of superconducting {\SRO} films was highly limited since its discovery \cite{YoshiharuPLD, RobinsonPLD}, and thus only bulk-based Josephson junctions, consisting of a {\SRO} bulk processed by focus ion beam (FIB) milling and a $s$-wave superconductor such as Pb \cite{SROJ_jin, SROJ_kidwingira, SROJ_nakagawa}, Sn \cite{SROJ_sumiyama}, and Nb \cite{SROJ_sumiyama, SROJ_nago, SROJ_saitoh1, SROJ_saitoh2, SROJ_anwar, SROJ_kashiwaya_arxiv} have long been studied. Such junctions have very large junction areas, compared to conventional film-based junctions. They are not a {\SRO}/{\SRO} Josephson junction, which facilitates simple analysis in theoretical models \cite{SROJ_p-p1, SROJ_p-p2, SROJ_p-p3, SROJ_p-p4} without considering possible mutual proximity effects in the {\SRO}/$s$-wave-superconductor junctions \cite{SROJ_s-p1, SROJ_s-p2, SROJ_s-p3, SROJ_s-p4}. It has also been reported that the the FIB process often damages the {\SRO} bulks, particularly forming other superconducting phase called the 3-K phase. However, recent development of molecular beam epitaxy growth of high-quality superconducting {\SRO} thin films \cite{UchidaMBE, TopoOxidereview, HariMBE}, as well as characterization of fundamental superconducting properties of the films \cite{Hc2_thinfilm}, is opening up a new route to study the unconventional superconductivity through film-based junction experiments.
 
Here we report fundamental properties of weak-link {\SRO}/{\SRO} Josephson junctions, fabricated from high-quality {\SRO} films with laser micro-patterning.  Planar tunnel junctions are other prototypical example, but which is rather difficult to make, because it needs epitaxial growth of high-quality {\SRO} films on an appropriate pin-hole free insulating barrier layer. The present film-based weak-link junctions, excluding possible proximity effects in the previous bulk-based junctions, are expected to enable more straight interpretation of characterization results. 


Superconducting single crystalline {\SRO} films with $c$-axis orientation were grown by molecular beam epitaxy with a thickness of $t=100$ nm on (001) (LaAlO$_{3}$)$_{0.3}$(SrAl$_{0.5}$Ta$_{0.5}$O$_{3}$)$_{0.7}$ (LSAT) substrates in a Veeco GEN10 system \cite{UchidaMBE}.  4N Sr and 3N5 Ru elemental fluxes were simultaneously supplied from a conventional Knudsen cell and a Telemark TT-6 electron beam evaporator, respectively. Molecular flux was calibrated using an INFICON quartz crystal microbalance system. The growth was performed at a substrate temperature of 900 $^{\circ}$C, regulated with a semiconductor-laser heating system, and with flowing pure $\mathrm{O}_{3}$ with a pressure of $1\times 10^{-6}$ Torr, supplied from a Meidensha Co. MPOG-104A1-R ozone generator. Figure 1(a) shows x-ray diffraction (XRD) $\theta$-2$\theta$ scan of a {\SRO} film used the junction fabrication. It shows only sharp (00$l$) {\SRO} peaks ($l$: even integer) up to (0014) and no discernible impurity peaks due to slightly excess supply of Ru. This indicates that single-crystalline and stoichiometric {\SRO} is almost homogeneously grown over the wide film area. Junctions with a narrow constriction were fabricated by selectively ablating a part of {\SRO} films, with the irradiation of a second harmonic Nd:YAG laser (532 nm wave length, 5 nsec duration) for 10 seconds at a repetition frequency of 30 Hz. The pulsed laser beam was focused down to 1 $\mu$m spot through optical slits and collective lenses and scanned at 250 nm steps to make the constriction with a width of $w=2$ $\mu$m. Aluminum wires for the four-terminal measurement were connected using a wire bonding machine. The junctions were cooled down to 60 mK in an Oxford Instruments Kelvinox MX100 $^3$He-$^4$He dilution refrigerator equipped with a superconducting magnet. Voltage across the constriction was measured with a Keithley 2182A nanovoltmeter while flowing a DC current along the $a$-axis with a Yokogawa 7651 programmable source.


Figure 1(b) shows temperature dependence of the in-plane resistivity of the {\SRO} film used for the junction fabrication. It exhibits a superconducting transition at a midpoint critical temperature of $T_{\mathrm{c, mid}} =1.2$ K with a width of $\Delta T =0.2$ K. These values are the best ones among reported unstrained films grown on the cubic LSAT substrate \cite{YoshiharuPLD, RobinsonPLD, UchidaMBE, Hc2_thinfilm}. The epitaxial strain from the LSAT substrate is almost biaxial and a change in the in-plane lattice parameters compared to bulk is further suppressed at low temperatures ($\varepsilon_{xx,yy} \sim +0.03$ \%) \cite{Hc2_thinfilm}, not requiring the consideration of uniaxial strain effect \cite{HariMBE, Hc2strain} in the present experiment. Figure 1(c) displays a {\SRO}/{\SRO} junction fabricated from this film, where two {\SRO} superconductor banks are connected through a narrow constriction. As confirmed in Fig. 1(d), the superconducting transition in the junction starts only slightly below the onset critical temperature of the original film. The laser fabrication process successfully avoids serious damage on the film, particularly formation of the 3-K phase as caused by the FIB process.

Figure 1(e) summarizes $I$-$V$ characteristics and its temperature dependence observed in the {\SRO}/{\SRO} junction \cite{supplement}. It shows a typical overdamped behavior of Josephson junctions, with $I_{\mathrm{c}} = 250$ $\mu$A and a normal state resistance of $R_{\mathrm{n}} = 32$ m$\Omega$ at the base temperature.  {\IcRn} product, expressed by the product of {\Ic} and {\Rn} as a measure of the superconducting coupling strength, is $8$ $\mu$V, which is comparable to the values previously reported for bulk-based {\SRO} junctions \cite{SROJ_saitoh1, SROJ_saitoh2, SROJ_kidwingira, SROJ_sumiyama, SROJ_kashiwaya_arxiv} but still lower than the reported superconducting gap amplitude of about $300$--$500$ $\mu$eV \cite{SRO_STMgap1, SRO_STMgap2}. On the other hand, the junction cross-sectional area in the present study is largely reduced to $A=wt=0.2$ $\mu$m$^2$, typically by two to four orders of magnitude compared to the previously reported bulk-based junctions \cite{SROJ_saitoh1, SROJ_saitoh2, SROJ_kidwingira, SROJ_anwar, SROJ_kamabara1, SROJ_nago, SROJ_kashiwaya_arxiv}. Namely, much lower normalized resistance $R_{\mathrm{n}}A$ of about $6\times10^{-11}$ $\Omega$cm$^2$ and higher critical current density $J_{\mathrm{c}}$ of about $1\times10^5$ A/cm$^2$ are achieved in the present Josephson junction, probably because of the all-{\SRO} structure avoiding interface problems. 

As shown in Fig. 2(a), {\Ic} shows clear modulations responding to the applied magnetic field. In thin films of type II superconductors, a characteristic length scale of the magnetic field distribution around a vortex is given not by the London penetration depth $\lambda$, but by the Pearl penetration depth $\Lambda=2\lambda^2/t$ especially in the thin limit ($t \ll \lambda$) \cite{textbook2}. Considering that $\lambda$ in {\SRO} bulks is about $200$ nm at the base temperature \cite{SRO_pd1_sans, SRO_pd2_usR, SRO_pd3_msi}, modulation period in the conventional Fraunhofer pattern is estimated between $\Phi_0/2w\Lambda=7$ Oe and $\Phi_0/2w\lambda=26$ Oe for the present junction. The fine oscillations appearing at intervals of about 15 Oe as seen in Figs. 2(b) and (c) are ascribed to this type of modulation. The oscillation pattern has been reproduced after wide sweeping of the magnetic field or recooling through the transition temperature, excluding possible domain contributions appearing as hysteresis in the bulk-based large junctions \cite{SROJ_kidwingira, SROJ_saitoh2, SROJ_kamabara1}. 

Figure 3(a) plots magnetic field dependence of {\Ic}, extracted from the map in Fig. 2(a). Here we conduct symmetrical analysis for the obtained $I_{\mathrm{c}} (H)$ curves by two types of operation. One is time-reversal operation, where both the current and field directions are inverted for {\Icn} taken at negative bias current, and then compared to {\Icp} taken at positive one (Fig. 3(b)). The other is current-reversal operation, where only the current direction is inverted for {\Icn} and then compared to {\Icp} (Fig. 3(c)). As confirmed in the comparison between $+I_{\mathrm{c,p}} (+H)$ and $-I_{\mathrm{c,n}} (-H)$ in Fig. 3(b), the observed oscillation patterns are almost unchanged by the time-reversal operation.  This is in contrast to the case of the current-reversal operation in Fig. 3(c), where such a fine agreement of the oscillation patterns is not discerned between $+I_{\mathrm{c,p}} (+H)$ and $-I_{\mathrm{c,n}} (+H)$. This mismatch is mainly ascribed to extrinsic self-field effect, where the trapped vortices impose a nonuniform contribution to the magnetic field. The oscillation curves {\Icp} and {\Icn} are then shifted in the opposite field direction, while time-reversal symmetry of the oscillation pattern is still preserved. In the case of chiral-$p$-wave/$s$-wave corner and planar Josephson junctions without time-reversal symmetry, time-reversal symmetry breaking oscillation patterns have been theoretically expected from a multiband model calculation including the spin-orbit interaction \cite{SROJ_s-p4, SROJ_kashiwaya_arxiv}. These patterns have been ascribed to symmetrically allowed cosine terms in the current-phase relation. However, the observed oscillation pattern is time-reversal invariant, and which cannot be explained by the mixed chiral domains, because such domain sizes become comparable to or even smaller than the in-plane coherence length \cite{SROreview1, Hc2_thinfilm}. On the other hand, the time-reversal invariant oscillations do not directly indicate that time-reversal symmetry is preserved in the {\SRO} system. In particular, so far it is not trivial that the above calculation based on the anomalous current-phase relation is simply applicable to other combinations of time-reversal symmetry breaking superconducting states such as chiral $d$-wave.

Finally, we discuss temperature dependence of the {\IcRn} product.  For weak-link junctions, the Kulik-Omelyanchuk model based on a point contact has been proposed, and temperature dependence of the {\IcRn} product has been theoretically derived for combination of $s$-wave superconductors \cite{IcRn_KO5}. On the other hand, the ideal magnitude relation between the in-plane coherence length and junction width ($\xi_{ab} \gg w$) is not satisfied in the present junction. As shown in Fig. 4, the observed temperature dependence seems deviated from the Kulik-Omelyanchuk relation in the present diffusive limit ($l \ll 2\lambda$) for the mean-free path $l$ of about 50 nm \cite{Hc2_thinfilm}. Tunnel-type junctions have been addressed by the Ambegaokar-Baratoff model, based on tunneling through a sandwiched insulator, but the observed temperature dependence is also further deviated from the relation derived for $s$-wave combination.  In chiral $p$-wave tunnel junctions, on the other hand, it has been theoretically predicted that {\IcRn} increases rapidly with approaching the base temperature, particularly for combination of the same chirality \cite{SROJ_p-p2}. However, such increase at low temperatures is not observed in the {\SRO}/{\SRO} Josephson junction. Although detailed temperature dependence may be dependent on multiband or anisotropic effects, the chiral $p$-wave state is unlikely also in terms of the temperature dependence.


In summary, we have fabricated the weak-link {\SRO}/{\SRO} Josephson junctions with enhanced critical current density, and investigated the fundamental properties including the responses to magnetic field and temperature changes. Observed magnetic field and temperature dependences in the present {\SRO}/{\SRO} Josephson junctions suggest that the chiral $p$-wave is unlikely for the superconducting symmetry, while general time-reversal symmetry breaking states cannot be ruled out. The results obtained on the first all-{\SRO} junction encourage further theoretical calculations considering other possible superconducting symmetry candidates \cite{SRO_NMR2, otherlist}. The potential of the film-based all-{\SRO} junctions has been also successfully demonstrated for further advancing studies by development of SQUID \cite{SROJ_s-p3, SROJ_s-p4} and grain boundary Josephson junctions \cite{J_p-FM1, J_p-FM3}, as well as theoretical analyses of the {\SRO}/{\SRO} junction properties. 


We acknowledge fruitful discussions with N. Nagaosa, R. Arita, J. Akimitsu, E. Zeldov, D. Manske, S. Ikegaya, K. Kanoda, Y. Iwasa, T. Shibauchi, T. Nojima, and S. Kittaka. This work was supported by Grant-in-Aids for Scientific Research (B) No. JP18H01866 and Scientific Research on Innovative Areas ``Topological Materials Science" No. JP16H00980 from MEXT, Japan and JST PRESTO No. JPMJPR18L2 and CREST Grant No. JPMJCR16F1, Japan.

\newpage

\begin{figure}
\begin{center}
\includegraphics*[width=11.0cm]{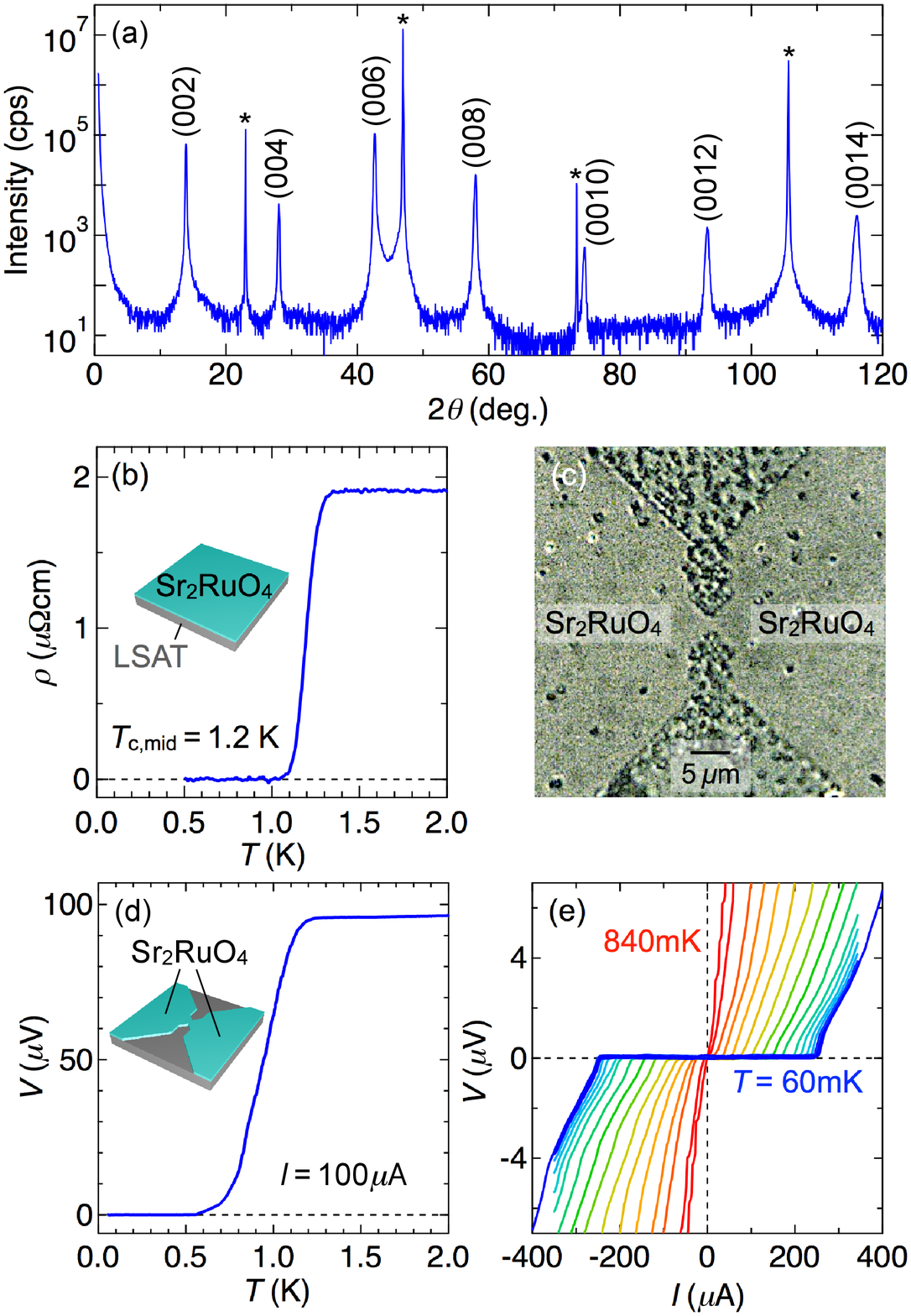}
\caption{Weak-link Josephson junction made from a {\SRO} thin film. (a) XRD $\theta$-2$\theta$ scan of the {\SRO} film used for fabricating the weak-link junctions. LSAT substrate peaks in the XRD scan are marked with an asterisk. (b) In-plane resistivity of the {\SRO} film before laser micro-patterning. It shows a sharp superconducting transition with a midpoint critical temperature of $T_{\mathrm{c, mid}} =1.2$ K. (c) Optical microscope image of a junction fabricated from the superconducting film, with the constriction width of $w=2$ $\mu$m. (d) Temperature dependence of voltage measured across the narrow constriction with feeding a current of $I=100$ $\mu$A. (e) Current-voltage characteristic of the weal-link {\SRO}/{\SRO} Josephson junction, taken at 60 mK and 90, 140, ..., 840 mK at intervals of 50 mK.
}
\end{center}
\end{figure}

\begin{figure}
\begin{center}
\includegraphics*[width=16.5cm]{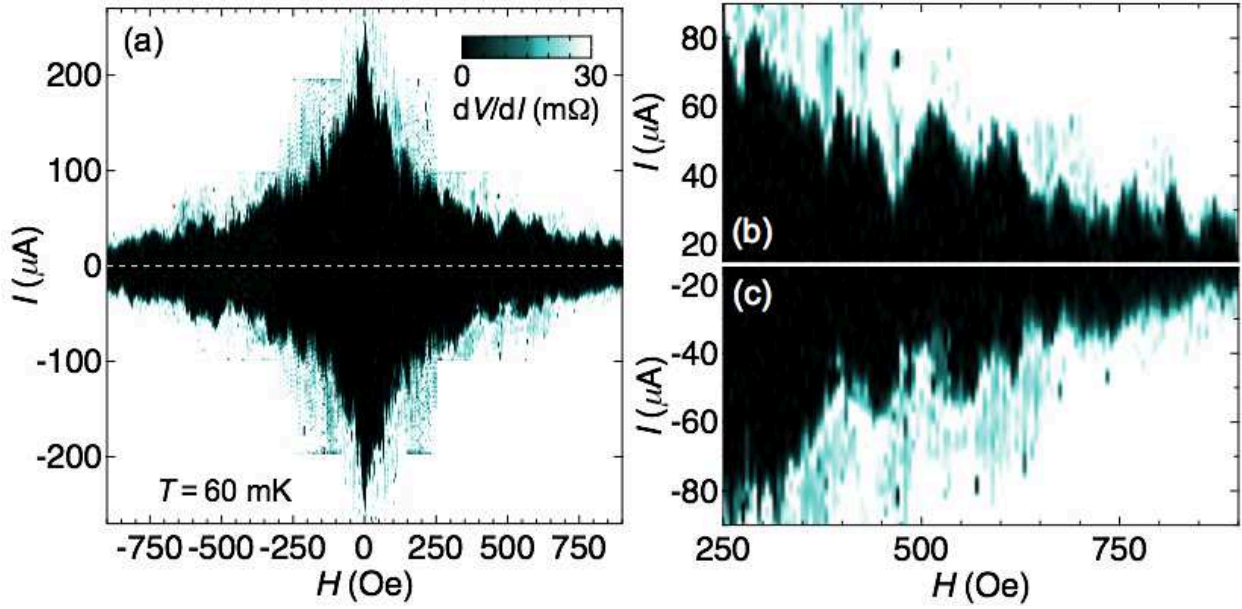}
\caption{Oscillation pattern observed in the {\SRO}/{\SRO} Josephson junction. (a) Differential resistance map taken as a function of current bias and out-of-plane magnetic field at the base temperature. (b) and (c) Magnified views of the oscillations for positive and negative bias.
}
\end{center}
\end{figure}

\begin{figure}
\begin{center}
\includegraphics*[width=16.5cm]{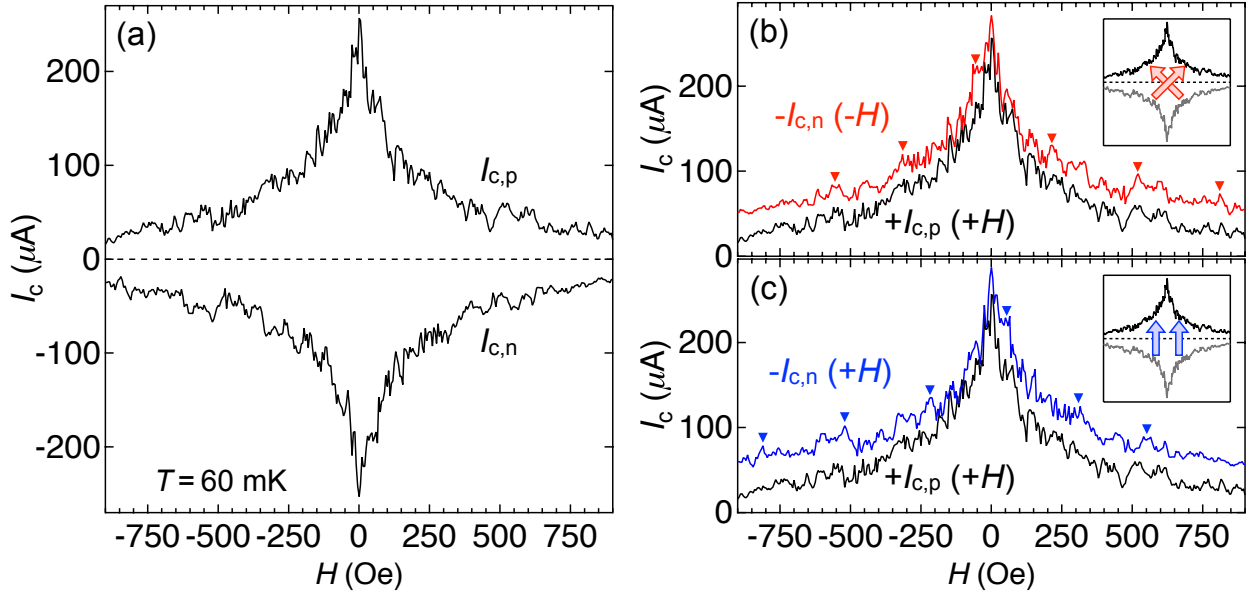}
\caption{Time-reversal invariant oscillations. (a) Magnetic field dependence of critical current extracted from Fig. 2(a). (b) and (c) Comparison of two oscillation curves {\Icp} and {\Icn} taken for positive and negative bias current. (b) {\Icp} is compared to $-I_{\mathrm{c,n}} (-H)$, where both the current and field directions are inverted for {\Icn} by time-reversal operation. (c) {\Icp} is compared to $-I_{\mathrm{c,n}} (+H)$, where only the current direction is inverted for {\Icn}.  Typical peak features showing clear match in (b) but not in (c) are indicated by a down-pointing triangle. 
}
\end{center}
\end{figure}

\begin{figure}
\begin{center}
\includegraphics*[width=7.5cm]{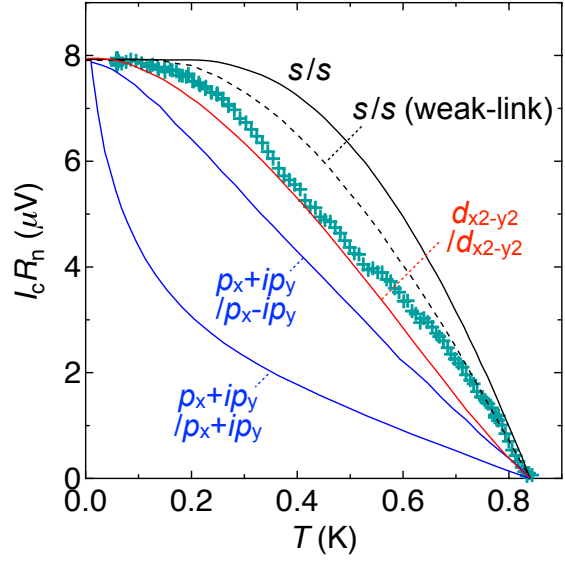}
\caption{Temperature dependence of {\IcRn} product is plotted with some fundamental theoretical results. The observation, denoted by plus symbols, shows some deviations from the Kulik-Omelyanchuk relation in diffusive weak-link junctions \cite{IcRn_KO5} as well as the Ambegaokar-Baratoff relation in tunnel junctions \cite{IcRn_s}, both of which are for $s$-wave/$s$-wave junctions. Tunnel junction results calculated for chiral $p$-wave combinations ({\pxyp}/{\pxym} and {\pxyp}/{\pxyp}) \cite{SROJ_p-p2} and a nodal $d$-wave combination ({\dxsys}/{\dxsys}) \cite{IcRn_dx2y2} are also compared. These calculations have been performed for the present in-plane configurations.
}
\end{center}
\end{figure}

\end{document}